\documentclass{elsart}
\usepackage[dvips]{graphicx}
\begin{document}
\begin{frontmatter}
\title{Clustering in a model with repulsive long-range interactions}

\author{Julien Barr\'e\thanksref{julien}}
and
\author{Thierry Dauxois\thanksref{thierry}}

\address{Laboratoire de Physique, UMR-CNRS 5672,
ENS Lyon, 46 All\'{e}e d'Italie, 69364 Lyon C\'{e}dex 07, France}

\author{Stefano Ruffo\thanksref{stefano}}
\address{Dipartimento di Energetica ``S. Stecco", 
Universit\'a di Firenze, Via S. Marta, 
3 I-50139, Firenze, Italy, INFM and INFN, Firenze}

\thanks[julien]{E-mail: jbarre@ens-lyon.fr}
\thanks[thierry]{E-mail: Thierry.Dauxois@ens-lyon.fr}
\thanks[stefano]{E-mail:ruffo@avanzi.de.unifi.it}

\begin{abstract}
A striking clustering phenomenon in the antiferromagnetic Hamiltonian
Mean-Field model has been previously reported. The numerically observed 
bicluster formation and stabilization is here fully explained by a non linear
analysis of the Vlasov equation.  
\end{abstract}
\end{frontmatter}
\vspace{-1.25truecm}
\section{Introduction}
\label{Introduction}
The Hamiltonian Mean Field model (HMF)~\cite{Antoni95} has attracted 
much attention in the recent years as a toy model to study the dynamics 
of systems with long-range interactions, and its relation to 
thermodynamics~\cite{Latora99}. Its Hamiltonian describes an
assembly of fully coupled rotors
\begin{equation}
H = \sum_{i=1}^{N}\frac{p_i^2}{2}+\frac{c}{2N}
\sum_{i,j=1}^{N}\cos(\theta_i-\theta_j)
\label{HMF}
\end{equation}
where $\theta_i$ is the angle of the rotor and $p_i$ its conjugate
angular momentum. Since we will use periodic boundary conditions, this
model can be alternatively viewed as representing particles moving on
a circle, whose positions are given by the $\theta_i$. In this paper
we consider the case in which the interaction among the particles is
repulsive (corresponding to the antiferromagnetic rotor model),
i.e. $c=1$.  As first noticed in Ref.~\cite{Antoni95}, this model has
a very interesting dynamical behaviour. In contrast with statistical
mechanics predictions, a bicluster forms at low energy for a special
but wide class of initial conditions (see Fig.~1 for the corresponding
density profile). For any initial spatial 
distribution of particles,
the bicluster develops in time from a homogeneous density state
as shown in Fig.~2, if the initial
velocity dispersion is weak. This clustering and the unexpected
dependence on initial conditions were studied in Ref.~\cite{DRH00} but
the phenomenon remained unexplained. In this paper we give an explanation
of the incipient formation of the bicluster based on a fully
analytical study of the low temperature solutions of the Vlasov
equation. We also show some connections of this problem with 
active transport in hydrodynamics~\cite{Castillo}, and the formation
of caustics in the flow of the Burgers equation.
\begin{figure}
%\label{densite1}
\centering{\includegraphics[width=0.75\textwidth,height=0.35\textheight]{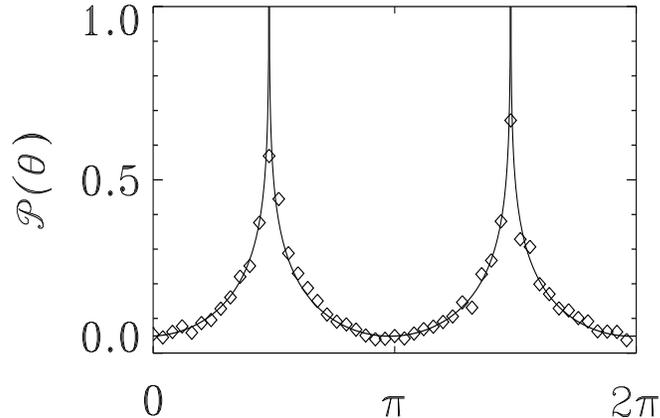}} 
\vspace{-1truecm}
\caption{Equilibrium density distribution for $N=10^4$ particles and the
energy per particle $U \simeq 10^{-5}$. The diamonds are numerical data and the 
solid curve is a fit taken from~\cite{DRH00}.}
\end{figure}
\begin{figure}
%\label{chevrons1}
%\centering{\includegraphics[width=0.75\textwidth,height=0.35\textheight]{instab2.ps}} 
\centering{\includegraphics[width=0.75\textwidth]{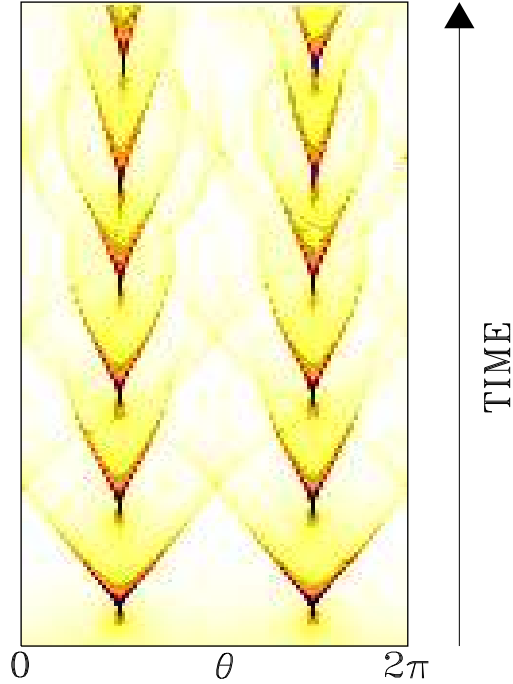}} 
\vspace{-0.5truecm}
\caption{Bicluster formation. Time is running in the upward direction and
the system is started with a constant particle density: the darker the grey, the bigger 
the density. One notices the very quick concentration of particles, followed
by the repeated appearance of ``chevrons'', shrinking as time increases.}
\end{figure}
\vspace{-1truecm}
\section{Hydrodynamical description at zero temperature}
We begin by performing first the thermodynamic  limit, $N \to \infty$, i.e. by writing the Vlasov equation
of Hamiltonian (\ref{HMF}). Calling $f(\theta,p,t)$ the one particle distribution 
function, we have
\begin{equation}
\frac{\partial f}{\partial t}+p \frac{\partial f}
{\partial \theta} -\frac{1}{2\pi}\int_{-\infty}^{+\infty} du 
\int_{0}^{2\pi} d\alpha \ f(\alpha,u,t) \sin(\theta-\alpha) 
\frac{\partial f}{\partial p} =  0
\end{equation}
We can define a density $\rho$, a velocity field $v$, and a velocity dispersion
$\sigma^2$ as follows
\vspace{-1truecm}
\begin{eqnarray*}
\rho(\theta,t) &=&  \int_{-\infty}^{+\infty}f(\theta,p,t)dp\\
\rho(\theta,t)v(\theta,t) &=&  \int_{-\infty}^{+\infty}pf(\theta,p,t)dp\\
\sigma^2(\theta,t)  &=& 
\frac{1}{\rho}\int_{-\infty}^{+\infty}p^2f(\theta,p,t)dp - v^2(\theta,t)~.
\end{eqnarray*}
The numerics shows that the bicluster appears when the velocity dispersion is
small. We will then neglect $\sigma^2$, which corresponds to a zero temperature 
approximation. A straightforward calculation reduces now the Vlasov
equation to the hydrodynamical equations for $\rho$ and $v$ 
\begin{eqnarray}
\frac{\partial \rho}{\partial t} + \frac{\partial (\rho v)}{\partial \theta} &=&  0 \\
\frac{\partial v}{\partial t} + v\frac{\partial v}{\partial \theta} &=&  
\frac{1}{2\pi} \int_{0}^{2\pi} d\alpha \ \rho(\alpha,t)\sin(\theta-\alpha)~.\label{euler}
\end{eqnarray}
The first equation accounts for the mass conservation and the second one is 
the Euler equation without pressure term. 
Hence, the dynamics of the model at zero temperature is mapped onto 
this active scalar advection problem~\cite{Castillo}, which we will 
now solve approximately at short time.
We first linearize the above equations, assuming the velocities are small and the
density almost uniform. We are left with
\begin{eqnarray*}
&&\frac{\partial \rho}{\partial t}  +  \frac{\partial v}{\partial \theta} =  0 \\
&&\frac{\partial v}{\partial t} = \frac{1}{2\pi}\int_0^{2\pi}\rho(\alpha,t)
\sin(\theta-\alpha)~,
\end{eqnarray*}
a system that we can solve by means of Fourier series. Assuming a vanishing
initial velocity field, we get
\begin{eqnarray}
&&\rho(\theta,t) =  \sqrt2 v_m\cos\theta\cos\omega t \nonumber \\
&&v(\theta,t)  =  v_m\sin\theta\sin\omega t~,
\label{sol1}
\end{eqnarray}
where $\omega=\sqrt 2/2$ is the plasma frequency. The amplitude $v_m$
and the longer wavelength components of $\rho$ in (\ref{sol1}) are fixed 
by the initial condition. 
This linear analysis describes very well the short time 
evolution of the system, a few plasma periods. Of course this is not sufficient 
to explain the bicluster formation and it is useful to carry out 
the non linear analysis: this analysis relies on the existence of two time-scales in the system. 
\begin{itemize}
\item The inverse of the plasma frequency, which is intrinsic and of order one. 

\item The timescale connected to the energy of the 
system of order $1/\sqrt U$. 
\end{itemize}

If the energy is sufficiently small, 
the two time-scales are very different, and it becomes  possible to use 
averaging methods. We introduce the long time-scale $\tau=\epsilon t$, 
with $\epsilon= v_m/\sqrt{2}$ ($v_m$ is directly related to the energy per
particle $U$), and we write
\begin{equation}
\label{vitesse} 
v(\theta,t) =  v_m\sin\theta\sin\omega t + \epsilon u(\theta,\tau) 
\end{equation}

We  have also to estimate the force in the non linear regime. As a zeroth
order approximation, we use the expression given by the linear analysis.
Since the force depends only on the first Fourier component of the density (due
to the special form of the Hamiltonian), this approximation amounts to 
suppose that the sinusoidal behaviour of this first component, found in the 
linear regime~(\ref{sol1}), remains valid in the non-linear regime.
This may appear very crude, but the numerics shows on the contrary that it is a quite
good approximation. This fact deserves a comment. Exploiting an analogy with the plasmas 
studied in Ref.~\cite{Elskens98}, our system may be seen as a bulk of particles 
interacting with waves sustained by the bulk itself. Here, there is one wave, 
materialized by the small oscillations in the density and the velocity field 
found in the linear approximation. Since at small energy the phase velocity 
of the wave (of order one like the plasma frequency) is much higher than 
the velocities of the bulk particles (of order $\sqrt U$), the wave has 
almost no interactions with these particles, and stands forever. This gives 
a qualitative explanation of the fact that the linear approximation for 
the force in  Euler equation (\ref{euler}) is so accurate. 
We introduce now expression (\ref{vitesse}) into equation (\ref{euler}). The
terms of first order in $\epsilon$ on the l.h.s. cancel the force in
the r.h.s.. The second
order terms gives
\begin{equation}
\frac{\partial u}{\partial \tau} 
+ u\frac{\partial u}{\partial \theta}
+2\sin\theta \cos\theta \sin^2\omega t
+\left(\sin\theta \frac{\partial u}{\partial \theta}
+u\cos\theta \right) \sqrt{2} \sin\omega t = 0 .
\end{equation}
Averaging over the short time scale, we get
\begin{equation}
\label{burgers} \frac{\partial u}{\partial \tau} 
+u\frac{\partial u}{\partial \theta}  =  -\frac{1}{2}\sin 2\theta\quad .
\end{equation}
This is a spatially forced Burgers equation without viscosity which describes
the motion of fluid particles in a potential given by
$V(\theta)=1/4\, \cos2\theta$. The double well shape of this potential is
responsible for the bicluster formation: particles will tend to spend more time
in the bottom of the wells. 
This equation may be solved using the method of characteristics, which presents a
great advantage: the characteristics are the Lagrangian trajectories of the 
Euler equation, so that they are an approximation 
to the particles trajectories of the real Hamiltonian system.
Since the characteristic equation is nothing but the equation of a pendulum
\begin{equation}
\frac{d^2\theta}{d\tau^2} + \frac{1}{2} \sin 2\theta =  0~,
\end{equation}
the Hamiltonian trajectories of the particles can be approximated by pendulum
trajectories (see Fig.~3).

We can also explain now the periodic recurrence and the shape of the 
``chevrons" (Fig.~2): they are zones of infinite density 
corresponding to the envelopes of the characteristics, so called caustics. 
Their approximate equation is 
\begin{equation}
\label{chevrons}
\theta \propto  \frac{\left( t-(2n-1)t_s \right)^{3/2}}{\sqrt{n}}
\end{equation}

\begin{figure}
%\label{carac}
\null\hskip 2truecm{\includegraphics[width=0.615\textwidth,height=0.286\textheight]{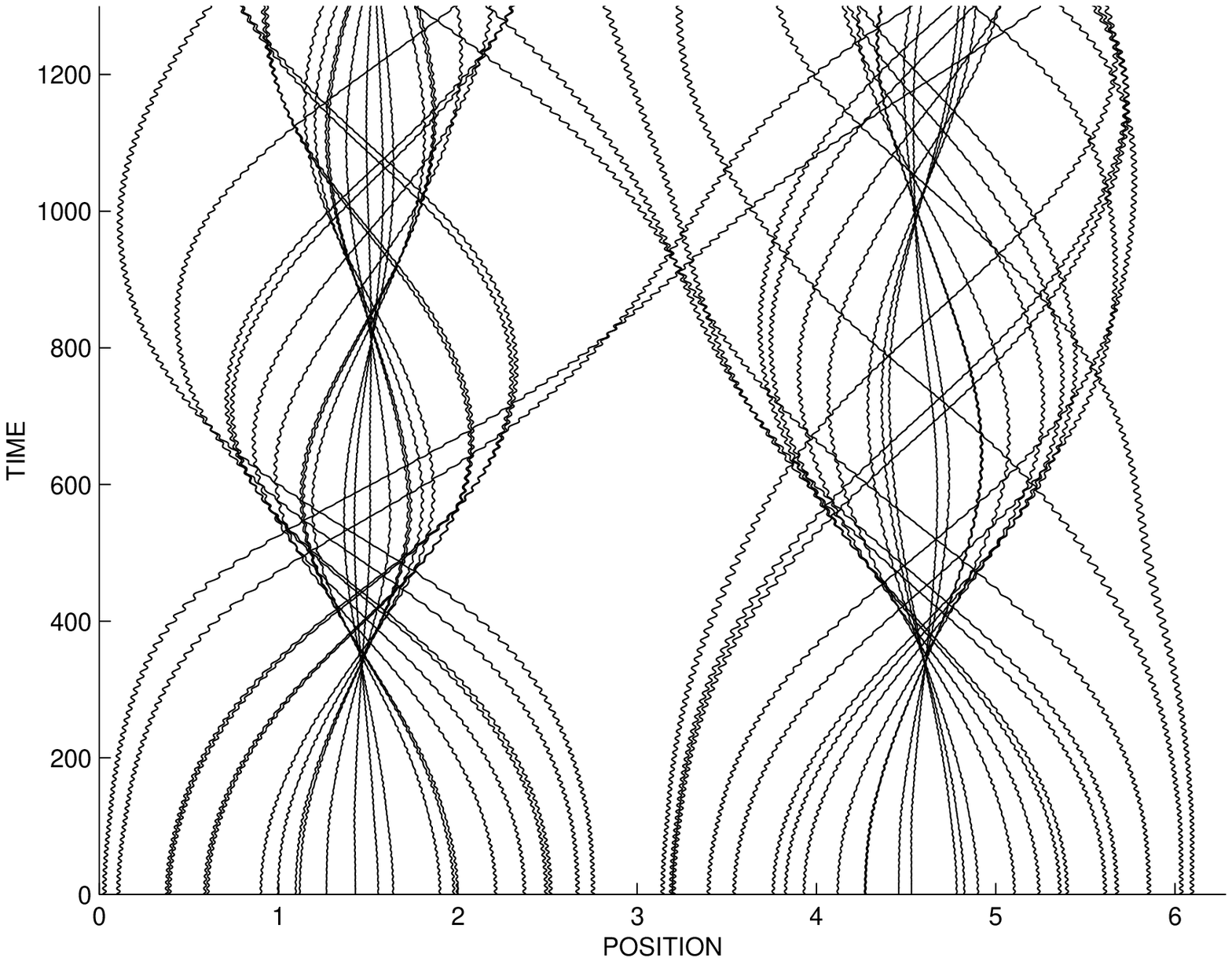}} 
\vspace{-2truecm}

\null\hskip 3truecm{\includegraphics[width=0.75\textwidth,height=0.35\textheight]{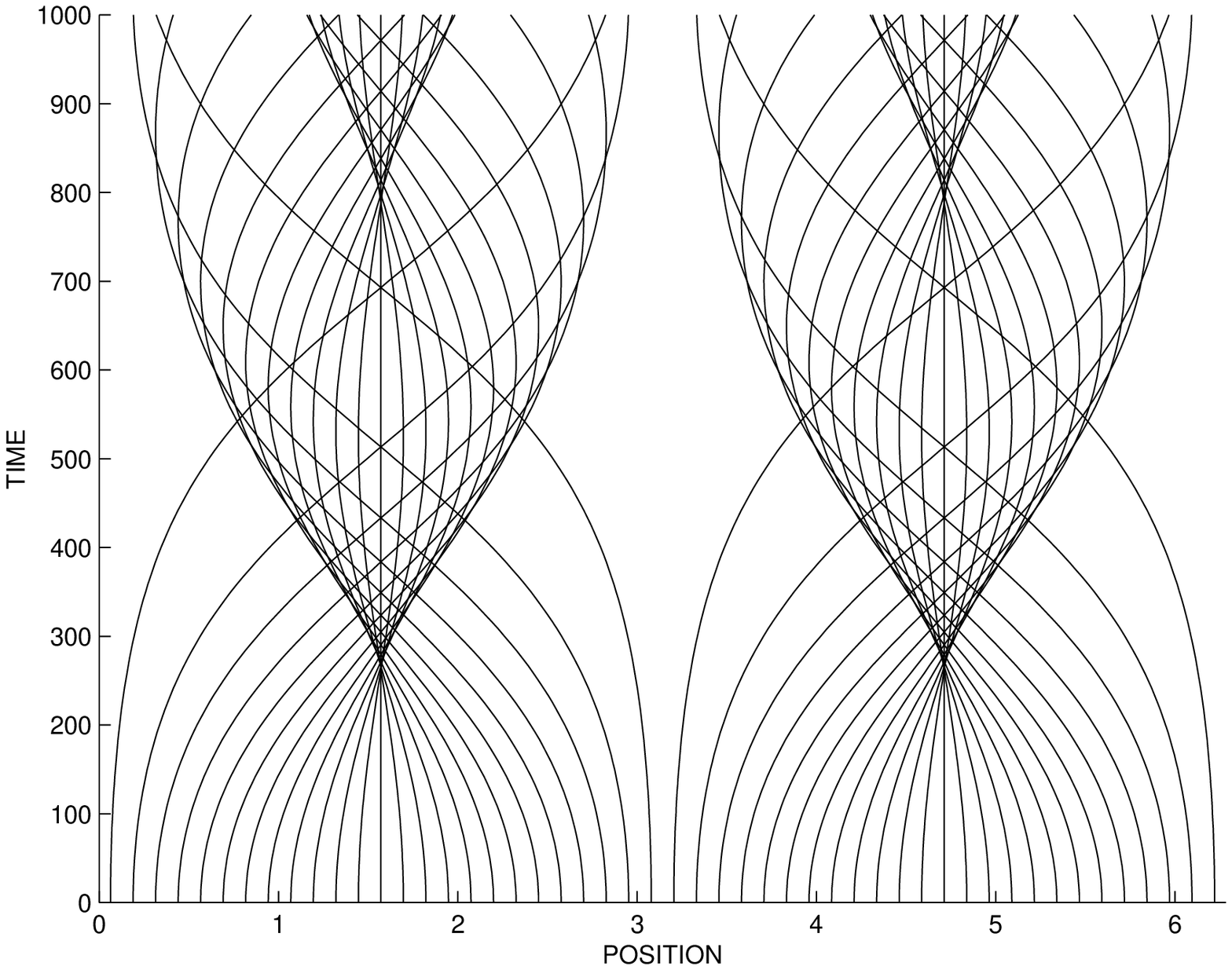}} 
\caption{Comparison between real trajectories of the particles (upper figure), 
and characteristics of equation (\ref{burgers}) (lower figure). One can see the 
two first appearances of the ``chevrons''. Two phenomena 
are not captured by the characteristics: the small and fast oscillations of the real 
trajectories which have been averaged out, and the presence of untrapped 
particles, close to the saddle-points of the effective $\cos2\theta$ potential.}
\end{figure}

where $n$ is the number counting successive ``chevron" appearences in time; 
the comparison with the numerics is drawn on Fig.~4. The $1/\sqrt n$ 
factor accounts for the shrinking of the ``chevrons''. 
Similar caustics are encountered in astrophysics, to explain the large scale 
structure of the universe: clusters and super clusters of galaxies are believed 
to be reminiscent of three dimensional caustics arising from the evolution of 
an initially slightly inhomogeneous plasma \cite{Zeldovich}.
\begin{figure}
%\label{chevrons_contour}
\centering{\includegraphics[width=0.6\textwidth,height=0.28\textheight,angle=90]
{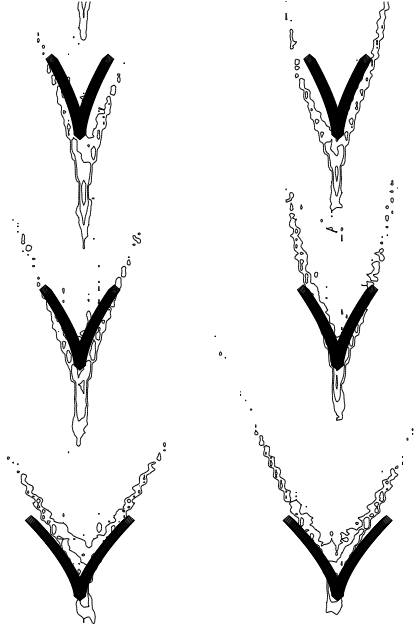}} 
\caption{Evolution of density for short times (level curves). The analytical formula 
for the ``chevrons'' (\ref{chevrons}) is superimposed (bold curves). 
$U\simeq 10^{-4}$ without adjustable parameter.}
\end{figure}
\section{Conclusions}
The surprising formation and stabilization of a bicluster in the 
HMF model~\cite{Antoni95} is now understood. The small collective 
oscillations of the bulk of particles create an effective double-well potential, 
in which the particles evolve.
This may be viewed as a nice example of active transport of the density
field~\cite{Castillo}, which can be studied analytically.
Lastly, we get an insight on the striking influence of initial conditions: 
when the initial thermal agitation is too strong, the description of the system
by density and velocity fields breaks down.
A peculiar behaviour of the trajectories evolving from initial conditions 
with a small velocity dispersion may be a somewhat general phenomenon 
for systems with long range interactions; indeed such features have
been very recently observed for a self-gravitating system in 1D~\cite{Konishi00}.
\begin{ack}
We thank J.L. Barrat, Y. Elskens, M.C. Firpo, D.H.E. Gross, P. Holdsworth,
R. Livi and S. Mac Namara for stimulating discussions.
This work is supported by the INFM-PAIS project {\it
Equilibrium and nonequilibrium dynamics in condensed matter}.
\end{ack}

\end{document}